\documentclass[journal]{vgtc}                
\ifpdf
  \pdfoutput=1\relax                   
  \usepackage{graphicx}                
  \DeclareGraphicsExtensions{.pdf,.png,.jpg,.jpeg} 
\else
  \usepackage{graphicx}                
  \DeclareGraphicsExtensions{.eps}     
\fi%

\graphicspath{{figures/}{pictures/}{images/}{./}} 

\usepackage{microtype}                 
\PassOptionsToPackage{warn}{textcomp}  
\usepackage{textcomp}                  
\usepackage{mathptmx}                  
\usepackage{times}                     
\usepackage{cite}                      
\usepackage{tabu}                      
\usepackage{booktabs}                  

\usepackage{soul}                      
\usepackage{wrapfig}                   
\usepackage{enumitem}				   
\usepackage[dvipsnames]{xcolor}        
\usepackage[final]{changes}

\definechangesauthor[name={Lewis Liu}, color=orange]{LL}

\definecolor{myblue}{RGB}{81, 167, 249}
\definecolor{myorange}{RGB}{243, 144, 25}
\definecolor{myred}{RGB}{239, 93, 87}
\definecolor{mypurple}{RGB}{179, 106, 226}
\definecolor{myyellow}{RGB}{245, 211, 40}
\definecolor{mygrey}{RGB}{166, 170, 169}

\onlineid{0}

\vgtccategory{Research}
\vgtcpapertype{application/design study}

\title{Supporting the Problem-Solving Loop: Designing Highly Interactive Optimisation Systems}

\author{Jie Liu, Tim Dwyer, Guido Tack, Samuel Gratzl, Kim Marriott}
\authorfooter{
\item
 Jie Liu, Tim Dwyer, Samuel Gratzl, Kim Marriott and Guido Tack are with Monash University.
 E-mail: jliu120@gmail.com, \{tim.dwyer3$\mid$guido.tack$\mid$kim.marriott\}@monash.edu, sam@sgratzl.com
}

\shortauthortitle{Liu \MakeLowercase{\textit{et al.}}: Supporting the Problem-Solving Loop: Designing Highly Interactive Optimisation Systems}

\abstract{Efficient optimisation algorithms have become important tools for finding high-quality solutions to hard, real-world problems such as production scheduling, timetabling, or vehicle routing. These algorithms are typically ``black boxes'' that work on mathematical models of the problem to solve. However, many problems are difficult to fully specify, and require a ``human in the loop'' who collaborates with the algorithm by refining the model and guiding the search to produce acceptable solutions. Recently, the Problem-Solving Loop was introduced as a high-level model of such interactive optimisation. Here, we present and evaluate nine \replaced{recommendations}{guidelines} for the design of interactive visualisation tools supporting the Problem-Solving Loop. They range from the choice of visual representation for solutions and constraints to the use of a solution gallery to support exploration of alternate solutions. We first examined the applicability of the \replaced{recommendations}{guidelines} by investigating how well they had been supported in previous interactive optimisation tools. We then evaluated the \replaced{recommendations}{guidelines} in the context of the vehicle routing problem with time windows (VRPTW). To do so we built a sophisticated interactive visual system for solving VRPTW that was informed by the \replaced{recommendations}{guidelines}. Ten participants then used this system to solve a variety of routing problems. We report on participant comments and interaction patterns with the tool. These showed the tool was regarded as highly usable and the results \added{generally} supported the usefulness of the underlying \replaced{recommendations}{guidelines}.} 


\CCScatlist{ 
 \CCScat{K.6.1}{Management of Computing and Information Systems}%
{Project and People Management}{Life Cycle};
 \CCScat{K.7.m}{The Computing Profession}{Miscellaneous}{Ethics}
}

\vgtcinsertpkg

\begin{document}


\firstsection{Introduction}

\maketitle

Automatic optimisation is increasingly used to find high-quality solutions to hard, real-world
\replaced{resource allocation and scheduling problems, such as timetabling \mbox{~\cite{DBLP:conf/cp/StolevikNRF11}}, logistics\mbox{~\cite{handbook_transportation}}, scheduling of scientific observations\mbox{~\cite{DBLP:journals/constraints/SimoninAHL15}}, or planning of medical procedures\mbox{~\cite{DBLP:conf/iccS/BettsMRTLEH15}}. The problem is first modelled mathematically and then a constrained optimisation solver is used to find the best---or at least a very good---solution to the problem\mbox{~\cite{nocedal2006numerical}}.}{problems in all areas of our lives, such as timetabling\mbox{~\cite{DBLP:conf/cp/StolevikNRF11}}, logistics\mbox{~\cite{handbook_transportation}}, scheduling of scientific observations\mbox{~\cite{DBLP:journals/constraints/SimoninAHL15}}, or planning of medical procedures\mbox{~\cite{DBLP:conf/iccS/BettsMRTLEH15}}. The aim of an optimisation system is to find the best---or at least a very good---solution to a decision problem by modelling it mathematically using constraints and objective functions, and then using a constrained optimisation solver to find a solution\mbox{~\cite{nocedal2006numerical}}.}

\replaced{The traditional view of the optimisation solver is that it is a ``black box'' which does not support user interaction.
However, the optimisation community has found that in practice many real-world problems require the user to actively engage in the solution process. 
One of the main reasons for such \emph{interactive optimisation} is that a mathematical model necessarily simplifies the real-world problem\mbox{~\cite{anderson2000human, cummings2012human}}. Interaction allows the user to bring their additional knowledge into the solution process and means that the tool is less brittle and can more flexibly adjust to unforeseen situations.
In addition, interaction builds the user's trust and confidence in the solver\mbox{~\cite{liusubmitted}}.}{Traditionally the optimisation solver is viewed as a ``black box'' and does not support user interaction.
However, there is a growing realisation within the optimisation community that many real-world problems require \emph{interactive optimisation} in which the user is actively engaged in the solution process. 
One of the main reasons for including a ``human in the loop'' is that a mathematical model necessarily simplifies the real-world problem and so may not capture all aspects of it\mbox{~\cite{anderson2000human, cummings2012human}}.  Involving the user allows them to bring their additional knowledge into the solution process and also allows the tool to more flexibly adjust to unforeseen situations.
Furthermore, interaction builds the user's trust and confidence in the solver\mbox{~\cite{liusubmitted}}.}

\replaced{The design of visual interfaces that support people to work effectively with optimisation systems can be seen as an application of Visual Analytics\mbox{~\cite{liu2018understanding}}.
Here we present and provide an initial evaluation of the first design recommendations for such interfaces.}{The design of effective visual interfaces for humans to control and explore optimisation can be seen as an application of visual analytics\mbox{~\cite{liu2018understanding}}.
Here we present and evaluate the first design guidelines for such visual interfaces.}
Our specific contributions are fourfold.
\begin{enumerate}[noitemsep,topsep=0pt,leftmargin=4mm]
\item Nine \replaced{recommendations}{guidelines} informed by the \emph{problem-solving loop}~\cite{liu2018understanding}. This loop is a theoretical framework for understanding the high-level user goals and tasks in interactive optimisation and is an analogue of the sense-making loop widely used to understand visual analytics. The \replaced{recommendations}{guidelines} cover visualisation of solutions and constraints, interaction with the solver, as well as comparison of solutions and provenance of solutions.
\item A review of representative interactive optimisation tools to determine how well these \replaced{recommendations}{guidelines} are supported by current tools. While some \replaced{recommendations}{guidelines} were well supported, most--such as showing a gallery of solutions, side-by-side comparison of solutions, showing solution provenance or feedback on the solution process--were not.
\item The design and implementation of an interactive optimisation tool that exemplifies all of the design \replaced{recommendations}{guidelines}. Our tool solves representative real-world problem with diverse constraints: the vehicle routing problem with time windows (VRPTW). 
\item A qualitative user study with 10 participants evaluating the interactive tool for solving VRPTW. The study explored: the usability of the tool; its flexibility in adjusting to new scenarios; the \replaced{tool's}{tools} support for our \replaced{recommendations}{guidelines}; and the usefulness of those \replaced{recommendations}{guidelines} as evidenced by the \replaced{tool's}{tools} use.
\end{enumerate}
This research informs the design of future optimisation tools. More broadly it informs the design of other kinds of AI-based decision-support tools, and visual analytics tools in general.


\section{Background}
\noindent \textbf{Interactive Optimisation:} 
\replaced{Optimisation uses computational techniques---such as integer programming, simulated annealing or constraint programming---to find the best (or at least a good) solution to a mathematical model of a decision problem.
The model specifies \emph{decision variables} for which the optimisation technique will find values, an \emph{objective function} to measure the quality of a solution, and \emph{constraints} that restrict the values that the decision variables can take.
Usually, we create a general model which is instantiated to a particular problem by assigning values to the model's \emph{parameters}.}{Optimisation solves real-world decision problems by building a mathematical model of the problem and using computational techniques---such as integer programming, simulated annealing or constraint programming---to try and find the best (or at least a good) solution to the model.
The model has \emph{decision variables} for which the optimisation technique will find values, an \emph{objective function} to measure the quality of a solution, and \emph{constraints} that restrict the values that the decision variables can take.
Usually the model is generic and is instantiated to a particular problem by assigning values to its \emph{parameters}.}

For example, in this paper we consider the Vehicle Routing Problem with Time Windows (VRPTW).
In this problem the model's parameters specify the number of vehicles available to visit customers, the time windows in which each customer must be visited, and the distance and travel time between customers and from each customer to the depot.
A solution to the problem assigns each customer to a vehicle, and determines the order of customer visits for each vehicle.
The constraints are that customers can only be visited in their time windows, each vehicle can only visit one customer at a time and must have time to travel between consecutive customers.
The objective is to minimise the total distance travelled by all vehicles.

\replaced{There is now a recognition by the optimisation community of the need in many applications for optimisation systems that are interactive rather than fully automatic ``black boxes'', for which the user simply provides values for the model's parameters and then waits for the system to compute a solution.
A recent survey paper\mbox{~\cite{meignan2015review}} clarifies the rationale for involving the user.
For many real-world problems it is unrealistic to fully model the problem. Allowing the user to modify constraints and adjust trade-offs in the objective function means that they can reduce the gap between the model and the real-world problem. It also provides flexibility as the user can modify the model to handle unexpected situations.
Other benefits of interaction are that it may allow the user to guide the optimisation algorithm to find a better solution and can help the user build an appropriate level of trust in the system.}{Optimisation is now routinely used to solve a wide variety of decision problems in many different application areas.
The standard approach has been to build a fully automatic system in which the solver is a ``black box'', the user simply provides the problem data and then waits for the system to compute a solution.
However, there is now a recognition by the optimisation community that in many applications there is a need to directly engage the user in the optimisation process. 
A recent survey paper\mbox{~\cite{meignan2015review}} clarifies the rationale for such \emph{interactive optimisation}.
For many real-world problems it is unrealistic to fully model the problem as it may have conflicting or uncertain objectives and constraints.
By modifying constraints and adjusting trade-offs between objectives the user can minimise the gap between the model and the real-world problem.
Allowing modification also provides flexibility as the model can be adjusted to handle unexpected constraints or objectives.
In addition, interaction allows the user to guide the optimisation algorithm to find a better solution and can help the user build an appropriate level of trust in the system.}


\replaced{Most research into interactive optimisation has focused on algorithms or applications.
For instance, the aforementioned survey\mbox{~\cite{meignan2015review}} almost totally ignores visualisation and interaction techniques, the user experience or user  evaluations.}{While interactive optimisation has been utilised in many different areas--vehicle routing\mbox{~\cite{anderson2000human, uugur2009interactive}}, design-related problems\mbox{~\cite{bailly2013menuoptimizer, brochu2010bayesian}}, and medical treatment\mbox{~\cite{thieke2007new, liu2018understanding}}--it has only recently been recognised as a topic worthy of study in its own right.
Most interactive optimisation papers have focused on algorithms or details of a particular application or tool.
For instance, the aforementioned survey\mbox{~\cite{meignan2015review}} focuses on the reasons and role of interactive visualisation, solver techniques and user preference gathering but almost totally ignores visualisation and interaction techniques, the user experience or user studies evaluating systems. 
} 
Two notable exceptions are an early review by Jones~\cite{jones1994visualization} of visualisation usage in optimisation and a survey by Miettinen~\cite{miettinen2014survey} of visualisations used for multi-criteria optimisation.
Other research, e.g. Goodwin et al.~\cite{goodwin2017constraint}, has investigated how to use visualisation to diagnose and refine optimisation models.
However, this is to support the initial development of the model by an expert in optimisation, not the subsequent use of the model by the end-user.

One area that has received some attention is whether interaction engenders trust in optimisation systems.
Liu et al.~\cite{liu2018understanding} found that participants gained trust through the use of interactive optimisation tools from a small qualitative study in brachytherapy.
And in two controlled studies Liu et al.~\cite{liusubmitted} found that feedback about the solution process led to over-trust, but allowing users to manipulate solutions led to better-calibrated trust. 

A number of user studies have evaluated the effectiveness of interactive optimisation systems~\cite{klau2010human, patten2007mechanical, thieke2007new, brochu2010bayesian, kapoor2010interactive, cummings2012human, bailly2013menuoptimizer, schumann2015interactive, holzinger2019interactive, belin2014interactive, cajot2019interactive, hakanen2011wastewater, caldas2016painting, schneeweiss2011fdconfig, coffey2013design, matejka2018dream, gajos2010automatically,dayama2020grids}.
In particular, Caldas and Santos~\cite{caldas2016painting} developed a user-guided interactive system for daylighting design and \replaced{found the proposed system could produce good-quality solutions.}{concluded the proposed system could produce good-quality solutions to meet desired performance requirements.}
Similarly, Matejka et al.~\cite{matejka2018dream} presented Dream Lens, an interactive visual analysis tool to explore and visualise generative design.
Another user study evaluated an interactive optimisation system developed by Coffey et al.~\cite{coffey2013design} that proposed a user-guided approach to allow users to directly navigate the simulation model and change the design, while the system adjusted the underlying parameters correspondingly.
\added{Dayama et al.~\cite{dayama2020grids} described an interactive optimisation system for layout design in which the tool suggested alternative layouts in an example gallery.}
Klau et al.~\cite{klau2010human} proposed an interactive optimisation system based on the human-guided simple search (HuGSS) framework to solve a particular kind of vehicle routing problem and reported that better solutions could be found by using the system.

However, to the best of our knowledge there has been no attempt to develop or evaluate general \replaced{recommendations}{guidelines} for the design of visualisation and interactions in interactive optimisation systems.


\noindent \textbf{Visual Analytics:}
Another essential tool for decision support is visual analytics.
This uses automatic analysis empowered by visualisation and interaction techniques to make sense of data~\cite{keim2010mastering}.
There are great similarities between visual analytics and interactive optimisation as both research fields aim to employ both humans and computers to solve complex real-world problems.
Indeed, in a sense interactive optimisation is a visual analytics task in which the user is trying to understand the space of possible solutions to a decision problem~\cite{liu2018understanding}.
Unlike interactive optimisation, usability and the creation of design guidelines has been an ongoing focus of visual analytics research.

There exist several models or frameworks describing high-level tasks and processes in visual analytics.
Pirolli and Card~\cite{pirolli2005sensemaking} proposed a sense-making loop which captured the processes and steps employed by analysts when analysing and making sense of information.
Keim et al.~\cite{keim2010mastering} presented the visual analytics process as a knowledge discovery framework describing typical interactions and tasks.
Ceneda et al.~\cite{ceneda2016characterizing} introduced a model, extended from Van Wijk's~\cite{van2006views} generic visualisation model, to describe the opportunities of using automatic guidance to support user analysis in visual analytics.
This model was further extended by Collins et al.~\cite{collins2018guidance} with a focus on improving the efficiency of the analytical process utilising different automatic guidance.
\added{Dudley et al.~\cite{dudley2018review} gives a workflow for interactive machine learning (ML).}


\added{\noindent \textbf{Guidelines:}
While currently there are no design guidelines for interactive optimisation systems, guidelines exist for other kinds of interactive applications. Endsley~\cite{endsley2017here} gives general guidelines for the design of autonomous and semi-autonomous systems.
\replaced{She}{He} highlights the importance of user involvement and control, recommending that tools support and assist decision making but do not make the final decision. Research into interactive ML highlights the need for meaningful explanations of system behaviour~\cite{kulesza2015principles}, the need to engage the user, provide effective data representations, and to exploit interactivity and provide rich interactions~\cite{dudley2018review}.
General design recommendations for AI systems emphasize the need for transparency in what the system can do and why it has made a particular decision or performed a particular action~\cite{amershi2019guidelines}. Nodalo et al.~\cite{nodalo2019building} also emphasizes transparency and the need to use simple, understandable visualisations. M{\"u}hlbacher et al.~\cite{muhlbacher2014opening} presents different models of interactive visualisation with analytic software and strategies for increasing user involvement.}

\noindent \textbf{Problem Solving Loop:} 
If we are to develop design \replaced{recommendations}{guidelines} for interactive optimisation tools it is crucial to understand the fundamental tasks and processes in interactive optimisation.
Hillier~\cite{hillier2012introduction} summarised the general steps in designing an automatic optimisation system, which they named the `design process'.
Shim~\cite{shim2002past} presented and explained the general decision-making process of decision support systems.
However, both Hillier and Shim only considered fully automatic optimisation rather than interactive optimisation.

\begin{figure}[tb]
 \centering
 \includegraphics[width=\columnwidth, trim=0.3cm 0cm 0.3cm 0cm,clip=true]{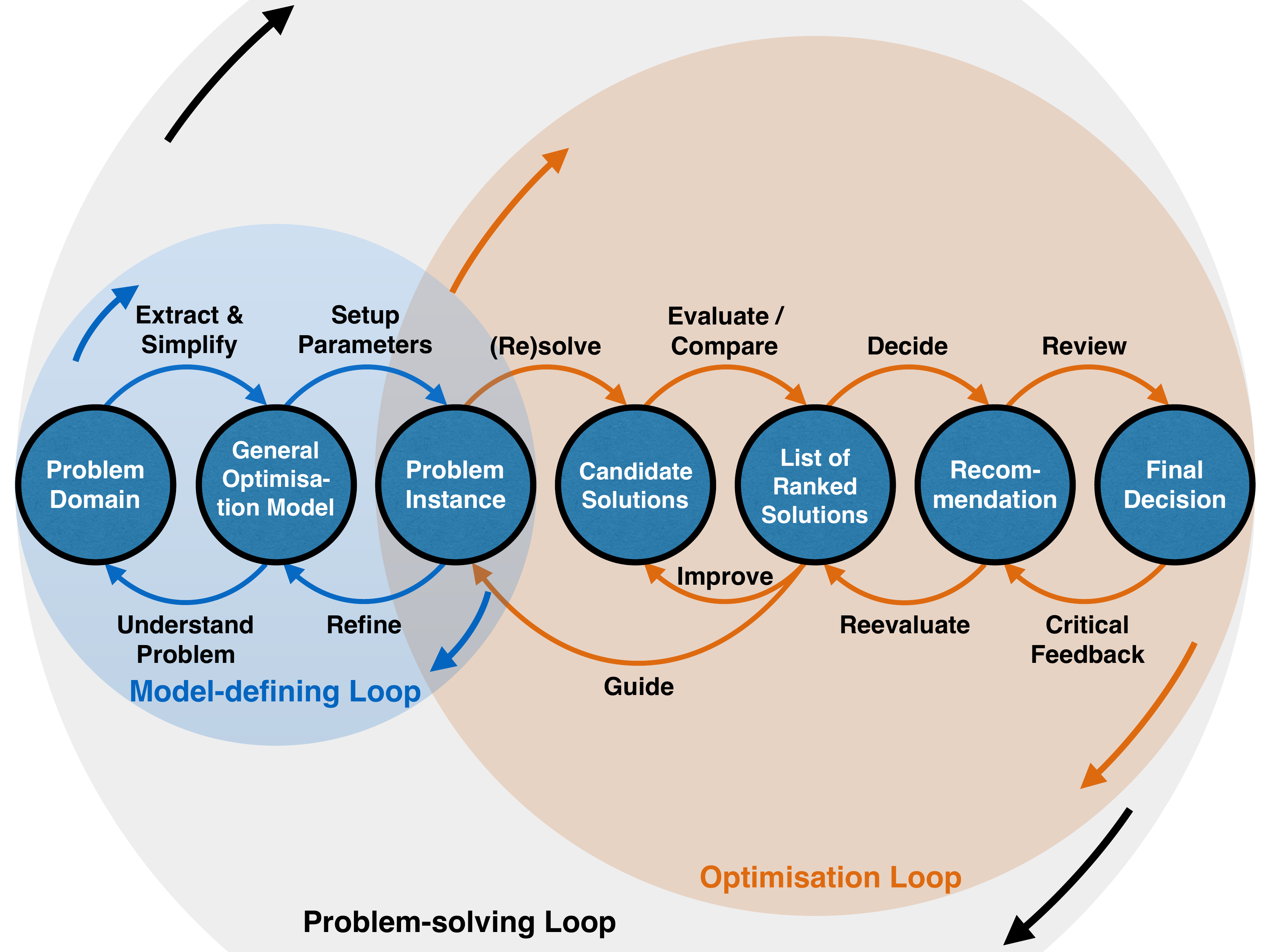}
 \vspace{-15pt}
 \caption{The problem-solving loop. Image from~\cite{liu2018understanding}.}
 \vspace{-10pt}
 \label{fig:solution-finding_loop}
\end{figure}

Recently, Liu et al.~\cite{liu2018understanding} proposed a theoretical framework called the \emph{problem-solving loop}, \added{see Fig.~\ref{fig:solution-finding_loop},} articulating the user goals and tasks in interactive optimisation. \added{It was based on feedback from health professionals on an interactive optimisation application in treatment planning for prostate cancer.}
\deleted{It is an analogue of Pirolli and Cards' sense-making loop\mbox{~\cite{pirolli2005sensemaking}} for visual analytics.}
The \replaced{problem-solving loop}{framework proposed by Liu et al.\mbox{~\cite{liu2018understanding}} (see Fig.~\ref{fig:solution-finding_loop})} has two main loops: the \emph{model-defining loop} and the \emph{optimisation loop}. 
The model-defining loop captures the process of creating an abstract mathematical model of the general problem, while the optimisation loop captures the interactive use of the model to solve a particular instance of the problem.
However, Liu et al.~\cite{liu2018understanding} did not evaluate this framework or use it as the basis for developing \replaced{design recommendations for}{guidelines for the design of} interactive optimisation systems.
This is the focus of this paper.
We now examine the loop more closely.


\vspace*{1mm}
\noindent\emph{Model Defining Loop:}
There are four tasks in the model-defining loop.
The two leftmost tasks \textit{Understand Problem} and \textit{Extract \& Simplify} capture the initial creation of the generic model by the (usually) expert optimisation modeller.
These tasks do not need to be supported by an interactive optimisation tool as the assumption is that they were completed when the tool was first created.
The other two tasks, however, should be supported in an interactive user interface. They are:
\begin{itemize}[noitemsep,topsep=0pt,parsep=0pt,partopsep=0pt,leftmargin=4mm]
\item \textit{Setup Parameters}: Determining the parameter values for the instance of the problem to be solved;
\item \textit{Refine}: Adding/removing/changing constraints and/or objectives in the optimisation model.
\end{itemize}

\noindent\emph{Optimisation Loop:}
The optimisation loop has eight tasks.
The rightmost four tasks do not need to be directly supported by the interactive optimisation tool as they are concerned with deciding on the best solution out of a ranked list of solutions and then the review of the solution by other stakeholders.
The \replaced{remaining}{following} four tasks, however, should be supported by the tool:
\begin{itemize}[noitemsep,topsep=0pt,parsep=0pt,partopsep=0pt,leftmargin=4mm]
\item \textit{(Re)solve}: Using the constraint solver to solve the generic model with problem-specific parameters to generate a pool of candidate solutions;
\item \textit{Evaluate/Compare}: Showing a specific single solution and its associated constraint(s) and objective(s) to support the solution evaluation process;
Or comparing multiple candidate solutions to give a ranked list of these solutions to better understand their trade-offs;
\item \textit{Improve}: Manipulating a solution to improve it;
\item \textit{Guide}: Guiding the optimisation solver by locking the satisfactory parts of a solution and leaving the remaining parts for the optimisation solver to change freely.
\end{itemize}

\section{Design \replaced{Recommendations}{Guidelines} for Interactive Optimisation}
\label{sec:guideline}
\replaced{Based on this analysis of the problem-solving loop we therefore recommend that an interactive optimisation tool should support the above six tasks. \replaced{This}{Such a tool} accords with Endsley's guidelines for autonomous systems~\cite{endsley2017here} as the task breakdown ensures that the user is in control of the solver's actions and remains responsible for choosing the final solution. It relies on a client-driven integration model in which the user via the visual interface controls the solver operation~\cite{muhlbacher2014opening}. We now give more detailed design recommendations informed by our literature review.}{Based on the problem-solving loop we have identified that an interactive optimisation tool should support the above six tasks.
Their consideration leads to the following guidelines.}\deleted{\mbox{\footnote{Note that these guidelines were included in Jie Liu's PhD thesis titled: ``Effective User Interfaces for Human-in-the-loop Optimisation'' but have not been published elsewhere.}}}

\vspace*{1mm}
\noindent \textbf{\emph{\replaced{Recommendation}{Guideline} 1}: \replaced{Provide appropriate}{Appropriate} visual representations of solutions \& constraints.}
Representations of solutions and constraints should, where possible, make use of visualisations currently used by domain experts so that they can understand and make sense of them.
Importantly, the visual representations of solutions and constraints should be tightly coupled.
Specifically, when presenting a solution the constraints also need to be represented in order to help users make sense of the solution and to \added{better} understand the solution space immediately around it.
It may look obvious to represent both solutions and constraints, but on occasion this has been overlooked by optimisation researchers ~\cite{deransart2000analysis}. 

\added{This recommendation accords with showing contextually relevant information~\cite{amershi2019guidelines}, providing effective data representations~\cite{dudley2018review} and use of understandable visualisations~\cite{nodalo2019building}.}
This \replaced{recommendation}{guideline} is central to supporting interactive \replaced{optimisation}{constraint-solving}. Without appropriate visual representations of solutions,
\replaced{\textit{Evaluating/Comparing} solutions and \textit{Guiding} the solver will be very difficult.}{\textit{Evaluating/Comparing} solutions as well as \textit{Guiding} optimisation solvers to find new solutions will be very difficult.}
Without appropriate visual representations of constraints, \replaced{\textit{Improving} a solution  and \textit{Refining} an optimisation model is difficult to achieve.}{\textit{Improving} a solution becomes more like a ``blind'' solution manipulation and \textit{Refining} an optimisation model is difficult to achieve.}

\vspace*{1mm}
\noindent \textbf{\emph{\replaced{Recommendation}{Guideline} 2}: \replaced{Support user modification of the optimisation model.}{Modifiable optimisation models.}}
The interactive tool should allow users to modify existing constraints, parameters and objective function. Furthermore they should be able to specify new constraints and add them to the optimisation model.

This \replaced{recommendation}{guideline} supports the \textit{Setup Parameters} and \textit{Refine} steps in the model-defining loop.
\added{It provides a semantically rich way of interacting with the solver~\cite{dudley2018review}}.
Importantly, being able to refine the optimisation model can produce more realistic and satisfying solutions by reducing the gap between the optimisation model and the real-world problem. It also supports dynamic optimisation if the optimisation problem changes.

\replaced{However, the ability to add constraints is limited by the solver's capabilities.
For instance, the user cannot add non-linear constraints to a linear solver. Furthermore, it can be difficult for non-experts in optimisation to understand abstract constraints.}{However, adding constraints on the fly will need to be limited for two reasons.
First, the new constraints and the type of optimisation solver need to match.
For instance, the user cannot add non-linear constraints to a problem solved by a linear solver without changing the solver.
Second, the capacity of the user to add constraints by him/herself may be limited, especially when the user does not have a background in optimisation.} Thus the kinds of constraints that can be added need to be understandable by a domain expert and appropriately represented (\replaced{Recommendation}{Guideline}~1).
Constraint acquisition might be a useful approach to elicit constraints from the user~\cite{de2018learning, arcangioli2016, beldiceanu2011constraint}.

\vspace*{1mm}
\noindent \textbf{\emph{\replaced{Recommendation}{Guideline} 3}: \replaced{Allow direct}{Direct} manipulation of solutions.}
Users should be able to directly manipulate a solution and make whatever changes they like to the values assigned to the decision variables and immediately see the impact on the objective function.

This \replaced{recommendation}{guideline} allows users to \textit{Improve} an existing solution.
Direct manipulation \replaced{and immediate feedback encourages user engagement~\cite{endsley2017here,dudley2018review} and supports efficient correction~\cite{amershi2019guidelines}.}{and interaction is a natural approach to change a solution.}
\deleted{Often, users are sceptical about solutions even if they look good.}
Being able to manipulate and change a solution allows users to \deleted{test the solution and} \textit{Evaluate} its quality in their own way.
With the support of appropriate representations of solutions and constraints, the user can explore the trade-offs between solutions, which is helpful for solution evaluation as well as perhaps finding a more satisfying solution. Prior research has found that direct manipulation allows the user to build an appropriate level of trust in the constraint solver~\cite{liusubmitted}.

\vspace*{1mm}
\noindent \textbf{\emph{\replaced{Recommendation}{Guideline} 4}: \replaced{Provide a gallery}{Gallery} of solutions.}
\replaced{Multiple solutions should be collected in a gallery.
This should provide an interface to explore the solutions and allow users to manually and automatically (re)order solutions based on, say, the value of the objective function.}{
Multiple solutions should be collected and represented in the form of a gallery.
This should provide the navigation interface for exploring the solutions.
The gallery interface should also allow users to manually and automatically (re)order solutions based on their preferences.}

The solution gallery supports \textit{Comparison} of multiple solutions, allowing users to create a \replaced{list of ranked solutions based on their preferences.}{ranked list of solutions}
\deleted{Manual re-ordering provides flexibility, while automatic re-ordering based on, e.g., the value of the objective function can speed up comparison and ranking.}
\replaced{The solution gallery is designed to encourage the user to develop and evaluate alternative solutions. This is particularly important in multi-criteria applications such as design where the objectives are conflicting and the trade-offs are unclear and  impossible to encode in a single objective function~\cite{dayama2020grids,todi2016sketchplore}.}{The solution gallery also gives the user an overview of all current solutions, which can be useful for a quick comparison of solutions or to navigate between solutions.}

\vspace*{1mm}
\noindent \textbf{\emph{\replaced{Recommendation}{Guideline} 5}: \replaced{Allow user controlled re-optimisation.}{User controlled re-optimise.}}
We believe it is necessary to provide users with the ability to \textit{re-solve} part of a solution. \deleted{based on their preferences.}
Often, users are satisfied with only some parts of a solution.
In that case, the system can support users by \replaced{allowing them}{providing the ability} to ``lock'' the values of variables in the good part of the solution, and \replaced{improve the remaining parts of the solution using the re-optimisation.}{improve the other parts of the by re-optimising.}
\added{Re-solving provides a semantically rich way of interacting with the solver~\cite{dudley2018review} by allowing the users to}\deleted{
In such a way, users can} \textit{Guide} the optimisation solver to explore the search space that they are interested in.
\deleted{Specifically,}Optimisation solvers should make minimal changes to the parts of a solution that are locked and apply any modifications to the unlocked parts without breaking any constraints.
This has the benefit of narrowing \deleted{down} the search space to the ``unlocked'' parts of the solution, so that optimisation solvers can find a better solution more quickly.
It also provides stability in the solving process, which helps to preserve the user’s mental model~\cite{purchase2006important, purchase2008extremes} \added{and accords with the guideline for AI systems to update and adapt cautiously~\cite{amershi2019guidelines}}.

\vspace*{1mm}
\noindent \textbf{\emph{\replaced{Recommendation}{Guideline} 6}: \replaced{Support}{Side-by-side} comparison of \deleted{two (or perhaps more)} solutions.}
To support better \textit{Comparison} of solutions the interface should allow two (or more) solutions to be viewed side-by-side.
\deleted{One important aspect is to allow users to verify details from either solution.}
\replaced{This allows the user}{Another is} to identify the similarities and differences between the solutions.
Visual cues may be used to highlight these.
\replaced{Side-by-side comparisons allows users to \textit{Evaluate/Compare} solutions in a more meaningful way. Solutions with equally-good values of the objective function may differ significantly.
Seeing and comparing the trade-offs between these solutions is required in order to find the best solution.}{Using side-by-side comparisons, users can more readily rank solutions and select the best one.
This is necessary because solutions with equally-good values of the objective function may differ significantly in aspects that may not be captured by the constraints or the objective.
Seeing and comparing the trade-offs between these solutions is necessary in order to find the best solution.
As a result, users can \textit{Evaluate/Compare} solutions in a more meaningful way.}

\vspace*{1mm}
\noindent \textbf{\emph{\replaced{Recommendation}{Guideline} 7}: \replaced{Generate diverse}{Generating different} solutions.}
We believe an ability to automatically generate a new solution that is significantly different to any of the solutions in the gallery, but which is still of high quality with respect to the objective function, will support the user finding better solutions\added{~\cite{dayama2020grids,koch2019may}}.
\replaced{It supports \textit{(Re)Solving} the optimisation problem to generate a pool of diverse candidate solutions and can help users better understand the problem by seeing possible trade-offs between different criteria. It helps the user to avoid ``decision biasing''~\cite{endsley2017here}.}{The purpose of finding different solutions is to better \textit{(Re)Solve} the optimisation problem.
This has the benefit of enriching the solution gallery so that the solutions have diverse characteristics.
Seeing different solutions can help users better understand the problem by knowing possible trade-offs between different criteria.} 

The last two \replaced{recommendations}{guidelines} are not directly motivated by the problem-solving loop but rather are general interface \replaced{recommendations}{guidelines}.

\vspace*{1mm}
\noindent \textbf{\emph{\replaced{Recommendation}{Guideline} 8}: \replaced{Provide feedback}{Feedback} on \added{the} solving process.}
Often, optimisation software gives little feedback about the solving process, or displays feedback at a very detailed, technical level. \replaced{This can be disquieting as finding a solution can take minutes, even hours or days. If possible provide the user with understandable feedback on what the solver is doing and how long they will have to wait. Doing so will increase trust in the optimisation software\mbox{~\cite{liusubmitted}}. Even something as simple as a progress bar can show users that the optimisation software is still properly functioning.}{
However, finding a solution can take minutes, even hours or days.
Keeping users waiting and staring at the screen with nothing happening is confusing and disquieting.
Even something as simple as a progress bar can show users that the optimisation software is still properly functioning, increasing their trust in the optimisation software\mbox{~\cite{liusubmitted}}.}

\added{Feedback on the solving process might include providing an explanation of the result. \emph{Explainability} of algorithms means to justify the decisions made by an algorithm in a form that is understandable by humans, or other algorithms. Many researchers are exploring explainable ML, e.g.~\cite{kulesza2015principles} as a way increasing transparency and user trust. Explainable optimisation is also a well established area~\cite{ocallaghan2005,junker2004}. However, in practice at present only some algorithms support it and the explanations can be difficult for non-expert users to understand.}

\vspace*{1mm}
\noindent \textbf{\emph{\replaced{Recommendation}{Guideline} 9}: Record solution provenance.}
As discussed in \emph{\replaced{Recommendation}{Guideline} 3} and \emph{7}, it is important to allow users to directly manipulate solutions and generate new solutions.
During this process it is likely that the users will wish to return to previous solutions.
This can be facilitated by storing the history of solution manipulations, and presenting the \textit{provenance} of each solution in terms of the steps that led to it.
\replaced{This encourages user interaction~\cite{endsley2017here}}{It also encourages the user to \textit{Manipulate} solutions} as there is no need to be afraid of making a mistake---they can always go back to a previous solution and start over again. In general, support for provenance is important in visual analytic tools~\cite{ragan2015characterizing}.
Another component of provenance is the ability to annotate solutions.
This can act as a memory aid for the person using the system and also as a way of communicating different aspects of the solution to stakeholders\added{~\cite{liu2018understanding}}.

\begin{figure}[tb]
 \centering
 \includegraphics[width=\linewidth, trim=0.1cm 19.2cm 14.7cm 0.1cm,clip=true]{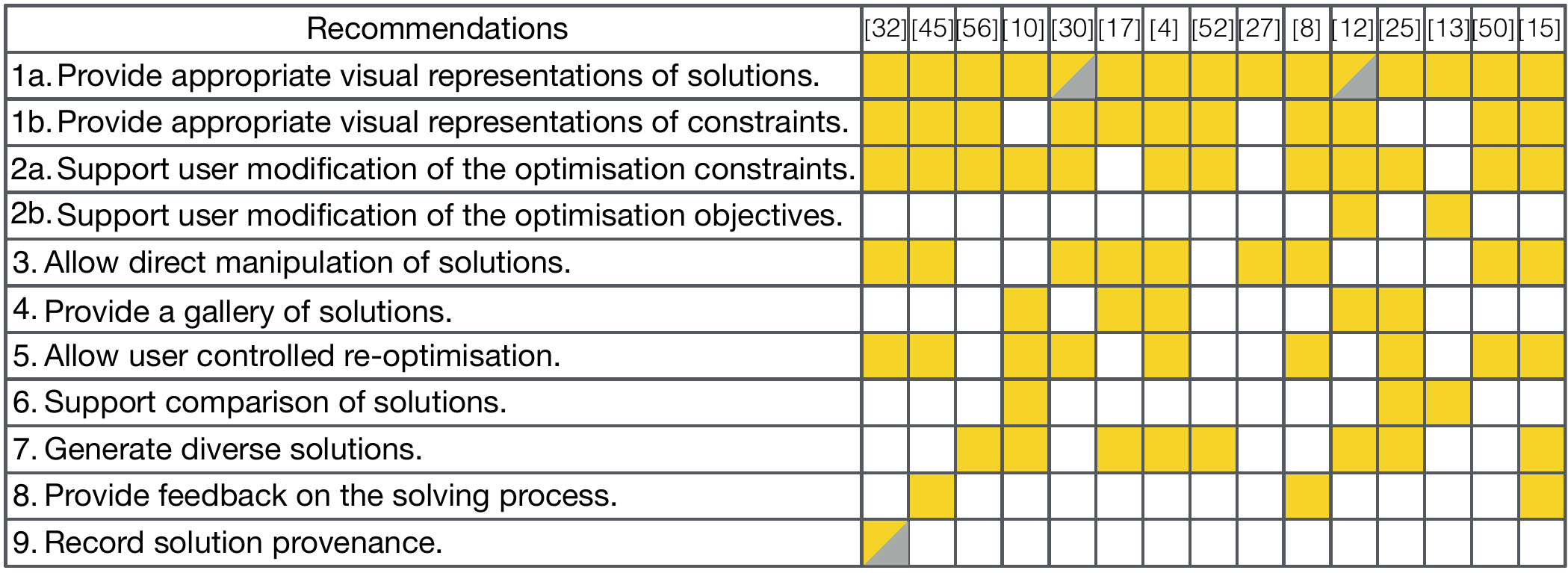}
 \vspace{-15pt}
 \caption{Reflections of the proposed design \replaced{recommendations}{guidelines} in representative systems from a variety of application domains. \textcolor{myyellow}{Yellow} cells represent full reflection. Partially \textcolor{mygrey}{shaded} cells represent partial reflection/exception. White cells indicate no reflection.}
 \vspace{-10pt}
 \label{fig:guideline-reflection-table}
\end{figure}

\subsection{Reflections of Design \replaced{Recommendations}{Guidelines} in Existing Systems}
In order to examine the usefulness of the nine \replaced{recommendations}{guidelines} proposed above, we will now examine (retrospectively) whether the \replaced{recommendations}{guidelines} are met by 15 representative systems from the literature~\cite{klau2010human, patten2007mechanical, thieke2007new, brochu2010bayesian, kapoor2010interactive, cummings2012human, bailly2013menuoptimizer, schumann2015interactive, holzinger2019interactive, belin2014interactive, cajot2019interactive, hakanen2011wastewater, caldas2016painting, schneeweiss2011fdconfig, coffey2013design}.
The examination is not exhaustive as there many interactive optimisation systems.
We chose 15 recent systems (selected from 2007 onward) to keep the reflection update-to-date and from different application domains to make the reflection as general as possible.
The findings are summarised in Fig.~\ref{fig:guideline-reflection-table}.

\begin{figure*}[t]
 \centering
 \includegraphics[width=\linewidth, trim=0cm 9.5cm 0cm 0mm,clip=true]{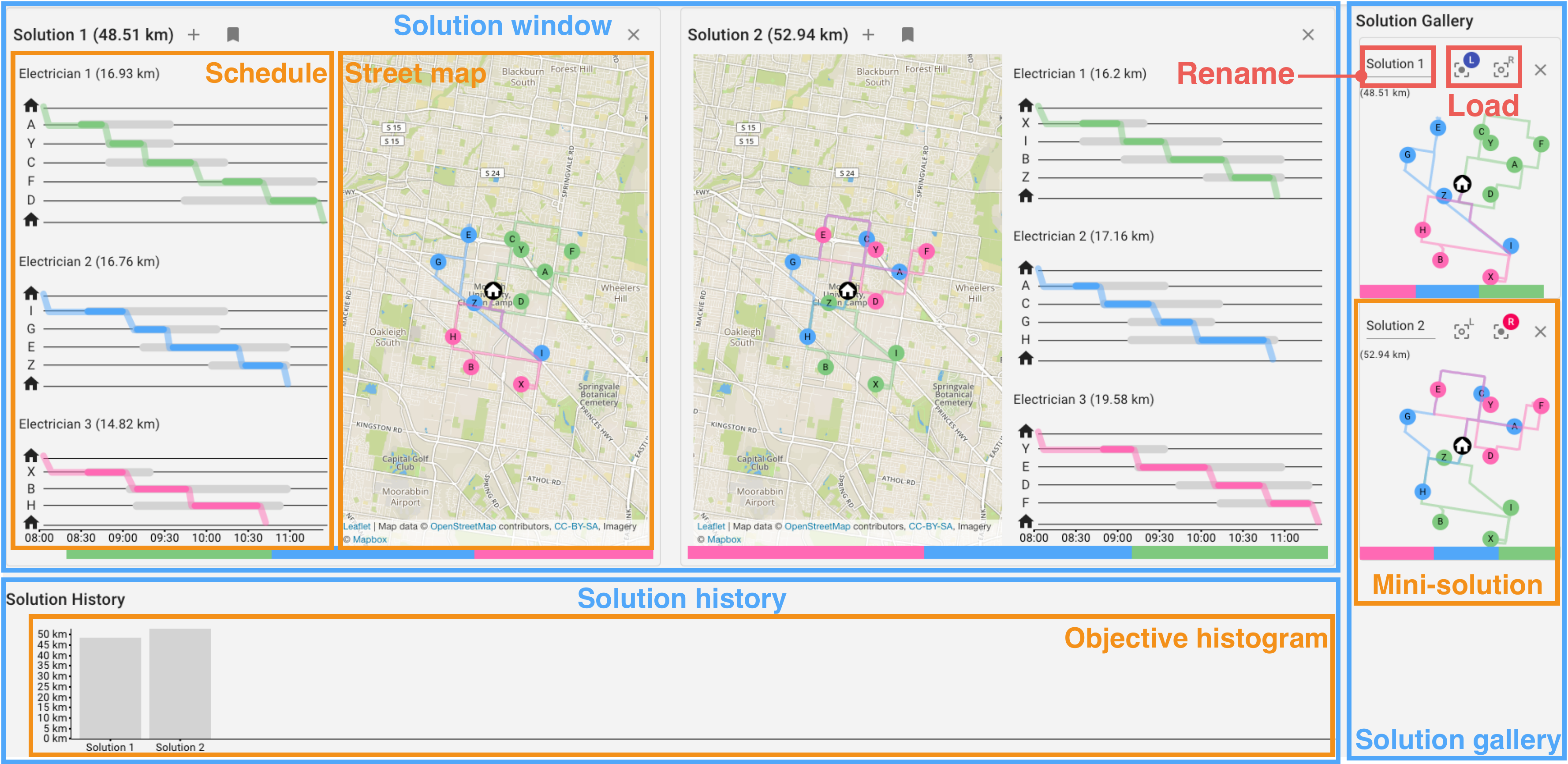}
 \vspace{-15pt}
 \caption{An overview of the interactive optimisation tool showing its three main components: \replaced{S}{s}olution window, \replaced{S}{s}olution history and \replaced{S}{s}olution gallery (highlighted in \textcolor{myblue}{blue}). Sub-components are highlighted in \textcolor{myorange}{orange}. Interactions are highlighted in \textcolor{myred}{red} (the rest of interactions in Fig.~\ref{fig:interface-solutionviewport}).}
 \vspace{-10pt}
 \label{fig:interface-overview}
\end{figure*}

\emph{\replaced{Recommendation}{Guideline} 1} is very well supported and is the most commonly supported \replaced{recommendation}{guideline}.
All of the examined systems have some kind of visual representation of solutions.
This is expected because a suitable visual representation of a solution is fundamental to making the results of the system understandable by the end-user. 

Most of these visual representations are domain-specific.
For example, in the survey paper of the Human-Guided Search project, Klau et al.~\cite{klau2010human} listed several interactive optimisation applications.
A node-link diagram was used in a graph layout problem.
A simplified map containing geographic locations of customers was used in a variation of Travelling Salesman Problem (TSP).
Similarly, Belin et al.~\cite{belin2014interactive} represented a city as a grid, in which different cells represented different elements such as roads and rivers.
However, there were two exceptions.
Unlike Belin's approach, Cajot et al.~\cite{cajot2019interactive} utilised parallel coordinates in an urban planning problem to facilitate stakeholders' decision making.
In machine learning, Kapoor et al.~\cite{kapoor2010interactive} used an adjacency matrix to represent the classification result. Both of these representations are unlikely to be understood by the average end-user.

The majority of the examined systems also have representations of constraints.
For instance, Cummings et al.~\cite{cummings2012human} represented impassable terrain such as big rocks as an area coloured in black in a path planning problem.
Bailly et al.~\cite{bailly2013menuoptimizer} drew a lock icon to enforce item placements without changes in a problem of application menu design.
However, not all systems visually represented constraints.
In the system developed by Brochu et al.~\cite{brochu2010bayesian}, user-defined constraints about model parameters for animation design were shown as a form using raw values.
    
\replaced{Recommendations}{Guidelines} 2, 3, 5 and 7 are supported by roughly half of the examined systems.
Specifically, for \emph{\replaced{Recommendation}{Guideline} 2}, some of these systems support constraint modification.
For example, Klau et al.~\cite{klau2010human} allowed users to modify the priority constraint associated with each customer in a VRP, which consequently affected re-optimisation.
Some systems utilise users' preference feedback to influence optimisation models and to guide optimisation solvers in producing more satisfactory solutions.
This appeared very common in multi-objective optimisation systems and other systems using genetic algorithms (GA).

For \emph{\replaced{Recommendation}{Guideline} 3}, more than half of the examined systems allowed users to make changes to solutions.
For instance, Patten et al.~\cite{patten2007mechanical} allowed users to directly move physical objects on a table-top interface to re-locate telephone towers when solving a telephone tower layout problem.


For \emph{\replaced{Recommendation}{Guideline} 5}, many examined systems provided re-optimisation.
Belin et al.~\cite{belin2014interactive} proposed an interactive urban city design tool which the optimisation solver simultaneously improves a city design while users are modifying the design. Most other tools however do not support such simultaneous collaboration and require the users and solvers to take turns to work on the problem~\cite{patten2007mechanical}.

For \emph{\replaced{Recommendation}{Guideline} 7}, half of the examined systems offered different solutions for the users to choose from.
Thieke et al.~\cite{thieke2007new} evaluated a medical treatment system which allows clinical professionals to explore multiple equally good treatment plans.
It becomes more important to generate and present different solutions to users to facilitate the trade-off comparisons between solutions, especially when there exist multiple conflicting objectives in an optimisation problem.
This led to better support of this \replaced{recommendation}{guideline} in multi-criteria objective problems.

\replaced{Recommendations}{Guidelines} 4, 6, 8 and 9 were poorly supported and only reflected in only a few of the examined systems.
This is perhaps because these \replaced{recommendations}{guidelines} support higher-level analysis and comparison of solutions rather than the more obvious interactions required to construct a solution.
Specifically, for \emph{\replaced{Recommendation}{Guideline} 4}, only Cummings et al.~\cite{cummings2012human} included a very basic solution gallery to facilitate comparison of different path plans.
For \emph{\replaced{Recommendation}{Guideline} 6}, Brochu et al.~\cite{brochu2010bayesian} provided four viewports with one solution each in a machine-learning-based optimisation system for animation design.
For \emph{\replaced{Recommendation}{Guideline} 8}, Coffey et al.~\cite{coffey2013design} simultaneously updated the model design when users were interacting and modifying the current design. 
Lastly, for \emph{\replaced{Recommendation}{Guideline} 9}, Klau et al.~\cite{klau2010human} mentioned that users were allowed to backtrack to previous solutions in the Human-Guided Search project, which provides a partial solution history.\footnote{It is possible that other systems also provided some kind of undo but that this was not mentioned in the system description.}

This suggests that while some of the \replaced{recommendations}{guidelines} are typically supported by interactive optimisation systems many are not.
The obvious question is therefore whether the proposed \replaced{recommendations}{guidelines} actually lead to more usable interactive optimisation tools.


\section{Interactive Optimisation System for VRPTW}
\label{sec:intopt}


In a second evaluation of the \replaced{recommendations}{guidelines} we used them to guide the development of an exemplar interactive optimisation tool and then conducted a user-study to evaluate the tool and the usefulness of the embodied \replaced{recommendations}{guidelines}. 

As our exemplar we chose to develop a tool for solving the Vehicle Routing Problem with Time Windows (VRPTW), a variation of the Vehicle Routing Problem (VRP). 
This is a practically important problem that is challenging to solve, but understandable by the general public.
The VRP problem is to schedule several \emph{trucks} to pick up goods from the \emph{home depot} and deliver them to the \emph{customers} who have ordered the goods.
This problem aims to minimise the total \emph{cost} of all \emph{truck routes}.
We take total distance travelled by all trucks (the sum of all \emph{truck route} lengths) as the \emph{cost}.
VRPTW extends VRP by requiring that a truck has to visit a customer to deliver a service within a certain \textit{time window}.
To make the problem even more challenging, we allowed a different \textit{service period} for each customer, which is the amount of time the vehicle requires at the customer site to deliver the service (e.g., unloading the goods).

\subsection{System Design}

We implemented our test system as a web application, with the back-end optimisation based on the MiniZinc~\cite{DBLP:conf/cp/NethercoteSBBDT07} system and the Gecode~\cite{gecode} constraint programming solver.

We spent considerable time refining and adjusting the web front-end to conform to the proposed \replaced{recommendations}{guidelines}. The tool contains three main components (see Fig.~\ref{fig:interface-overview}), as follows.

\noindent \textbf{A) Solution window:} 
In this window solutions may be explored in detail and interactively modified.
We designed the solution window to present either a single solution or two solutions shown side-by-side as recommended by \emph{\replaced{Recommendation}{Guideline} 6: \replaced{Support comparison of solutions.}{Side-by-side comparisons of two (or perhaps more) solutions.}}
This supports participants in comparing different solutions, but it also allows participants to compare a single solution before and after re-optimisation to see what has been changed. 
Each solution in the solution window is shown in its own \textit{solution viewport}, see Fig.~\ref{fig:interface-overview}.
Each solution viewport comprises two panels.\footnote{The representation of solutions is similar to that used in~\cite{liusubmitted}.
However, the system used in~\cite{liusubmitted} was a minimal implementation designed only to evaluate the effect of solver feedback and solution manipulation on trust and so provided limited support for interaction and no support for most \replaced{recommendations}{guidelines}.} 
    
\begin{figure}[t]
 \centering
 \includegraphics[width=\linewidth, trim=1.5cm 4.3cm 4.5cm 0mm,clip=true]{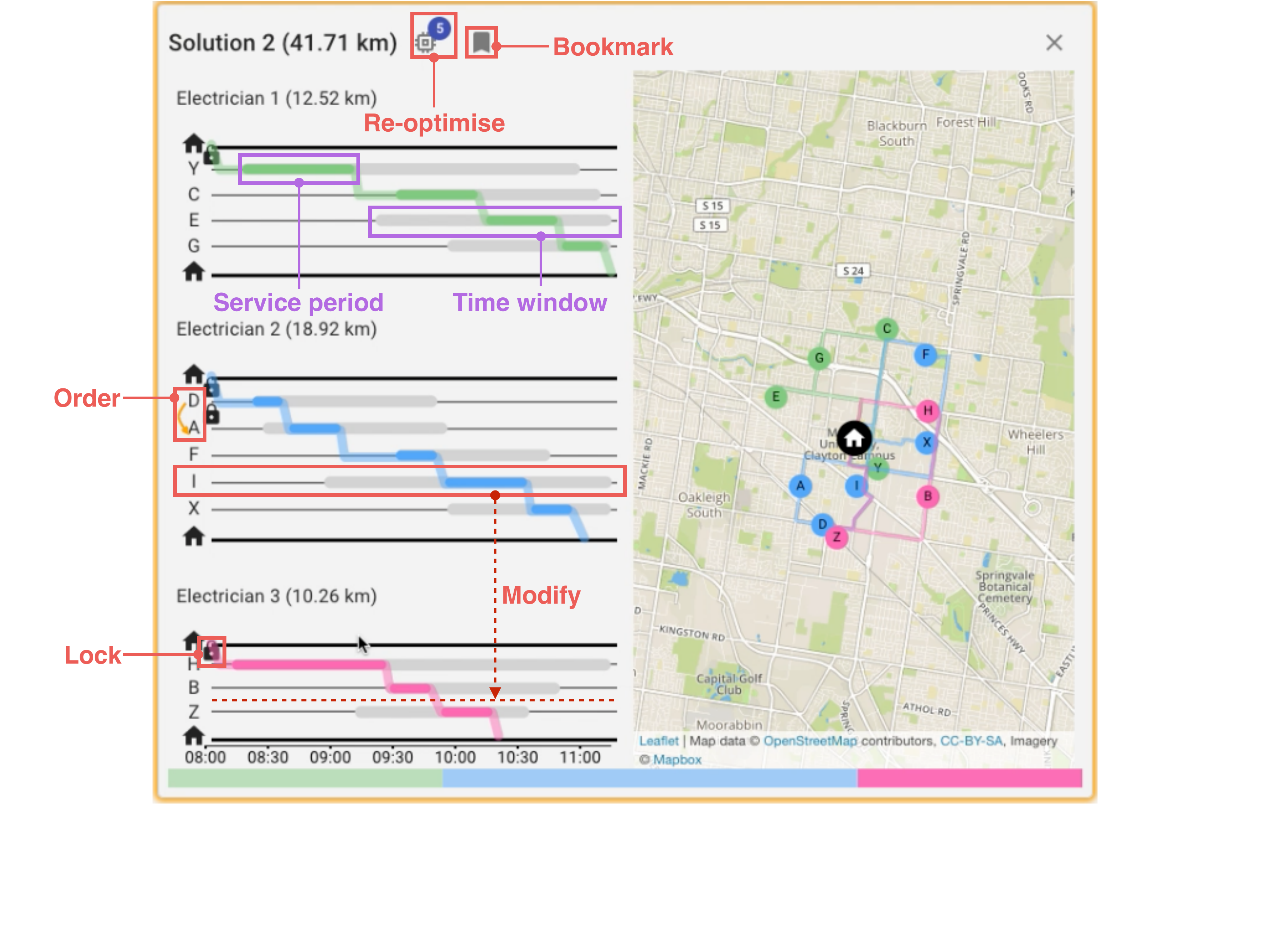}
 \vspace{-15pt}
 \caption{A detailed view of the \textit{solution viewport} in the solution window including the street map and the schedule. Interactions are highlighted in \textcolor{myred}{red}. Constraints are marked in \textcolor{mypurple}{purple}.}
 \vspace{-10pt}
 \label{fig:interface-solutionviewport}
\end{figure}

\noindent The \textbf{street map panel} (Fig.\ \ref{fig:interface-solutionviewport}-right) displays the geographic information about the trucks' \emph{home depot}, \emph{customers} (labelled A--Z) and \emph{truck routes}--each truck's route represented in a different colour.

We first considered a node-link diagram to display a solution with nodes representing customers (at their geographic locations) and a straight-line link representing the truck route travelled between customers.
While these diagrams were easy to draw, and reflect the underlying mathematical model used by the constraint solver, we felt that they violated \emph{\replaced{Recommendation}{Guideline} 1: \replaced{Provide appropriate}{Appropriate} visual representation of solutions}.
In reality, \replaced{while a node-link diagram may represent the accurate distance or time between nodes,}{} the route a truck travels from one location to another can be complex, with many turns at intersections and roundabouts.

We found a better representation using OpenStreetMap (OSM)\footnote{\url{https://www.openstreetmap.org}} for map display and Open Source Routing Machine (OSRM)\footnote{\url{https://project-osrm.org}} for realistic routing in road networks, together with Leaflet\footnote{\url{https://www.leafletjs.com}}, a JavaScript map annotation library, to create the street map representation for our tool.
Now, the street map view provides an unambiguous visual context of a VRPTW and the routes are represented \added{both} realistically and accurately\replaced{, being faithful to the underlying mathematical model}{}.
    
\noindent The \textbf{schedule panel} (Fig.\ \ref{fig:interface-solutionviewport}-left) shows the time window of each customer and the order of service delivery of each truck.
Marey diagrams~\cite{tufte2014visual} are used in transport schedules to represent locations of vehicles and the transport routes, which fits well in the VRPTW context.
In Marey diagrams time is showed vertically and locations are displayed horizontally.
However, to better suit the aspect ratio of the panel we chose to swap these. This also has the advantage of making customer names more visually salient than the time of delivery, reflecting their relative importance.

Thus a schedule has each customer represented by a horizontal line and the horizontal extent of the lines indicates the total time required for the schedule.
Customer lines are grouped by the truck which services them.
A sequence of line segments in the colour of the truck as used on the map, passes through the customer lines, indicating the truck's progress in making deliveries to customers.
Time windows for each customer are represented by a grey rectangle with the position of the left side of this rectangle indicating the start of the window, and the right side indicating the end of the window.
The other essential constraint is the service period, i.e.\ the time required by a truck to deliver a service.
We used another rectangle in the colour of the truck to represent a service period.
The rectangular representations of the time window and the service period were chosen for two reasons:
First, axis alignment facilitates quantitative comparisons of a service period and a time window of a single customer and comparisons of time windows of multiple customers.
Second, it visually distinguishes valid from invalid situations.
In a valid situation, a service period stays within the range of a time window resulting in the service period rectangle completely overlapped on top of the time window rectangle.
Whereas in an invalid situation, a service period goes beyond the close of a time window resulting in a possible partial overlapping between the service period rectangle and the time window rectangle.
Thus this representation conforms to the second part of \emph{\replaced{Recommendation}{Guideline} 1: \replaced{Provide appropriate}{Appropriate} visual representation of constraints}.

\noindent \textbf{B) Solution history:} keeps tracks of the participant's interaction history.
We designed a histogram to record all solutions and their provenances according to the \emph{\replaced{Recommendation}{Guideline} 9: Record solution provenance}.
    
\noindent \textbf{Objective histogram} represents each solution as a rectangular bar whose height indicates the value of the solution's objective function (total distance travelled).
Whenever constraints of a solution change or the solution is re-optimised, we draw a new bar for the new solution.  
We draw an arc between two solutions if the solution on the right is derived from that on the left.
Selecting a history bar brings the details of the corresponding solution into the solution window.
    
\noindent \textbf{C) Solution gallery:} supports participants managing solutions so as to finally select the best solution out of many, following \emph{\replaced{Recommendation}{Guideline} 4: \replaced{Provide a gallery}{Gallery} of solutions}.
The gallery supports annotations allowing participants to edit solution names (\emph{\replaced{Recommendation}{Guideline} 9: Record solution provenance}).

\noindent \textbf{Mini-solution} provides a simplified view of a solution by showing vehicle routes without the street map and including the objective value and the solution name.

The tool initialises the solution history and gallery with several different solutions to give users flexibility to choose between different starting solutions.
The solutions are chosen to be diverse.
We measured diversity between solutions using a diversity matrix, which calculated the routing sequence differences between one solution and all the others.
We select solutions with maximum diversity between the new computed solution and any previously computed solutions.
Providing such a selection of starting points meets \emph{\replaced{Recommendation}{Guideline} 7: \replaced{Generate diverse}{Generating different} solutions}.

\subsection{System Interaction}
Interactions fall into two groups: \textbf{optimisation-model interaction} and \textbf{problem-solving interaction}.
There are two optimisation-model interactions:

\noindent \textbf{Lock} allows the user to fix a specific customer to a specific truck.
A lock is drawn beside the customer to represent this constraint. This allows us to specify, for instance, that a customer requires the use of a particular truck. Locking a customer to a truck does not constrain the time of service by the truck.
        
\noindent \textbf{Order} fixes the relative order in which two customers of the same truck are to be serviced.
A directed arc line is drawn from one customer to another to represent this constraint.
It is useful in pick-up and delivery problems, where the goods to be delivered to a customer may be located at another customer's location.
Thus, the goods from the second customer need to be picked up first and then delivered to the first customer.


These constraints, supported by appropriate visual representations (\emph{\replaced{Recommendation}{Guideline} 1: \replaced{Provide appropriate}{Appropriate} visual representation of constraints}), give users the flexibility to modify the optimisation models, \emph{\replaced{Recommendation}{Guideline} 2: \replaced{Support user modification of the optimisation model.}{Modifiable optimisation models.}}

There are five problem-solving interactions. These allow the user to create new solutions.
\noindent \textbf{Modify} allows users to re-assign customers within the same truck or between two trucks via drag-and-drop.
This conforms to \emph{\replaced{Recommendation}{Guideline} 3: \replaced{Allow direct}{Direct} manipulation of solutions}.

\noindent \textbf{Re-optimise} allows the user to collaboratively solve an optimisation problem with the optimisation solver with the control of when and what to re-optimise.
This is a reflection of the \emph{\replaced{Recommendation}{Guideline} 5: \replaced{Allow user controlled re-optimisation.}{User controlled re-optimise.}}
In particular, users can freeze a part or multiple parts of a solution using the optimisation-model interactions described above, and then re-optimise the solution by clicking the re-optimise button located at the very top of the solution viewport within the solution window.

Before the optimisation solver spits out a new solution, the re-optimise button is replaced by a spinning wheel to indicate the ongoing re-optimisation process.
This is in accord with \emph{\replaced{Recommendation}{Guideline} 8: \replaced{Provide feedback}{Feedback} on \added{the} solving process}.

\noindent \textbf{Load} allows users to choose a solution from either the solution history or the solution gallery and to display the solution on either the left or the right solution viewport in the solution window by clicking the load button when hovering over the target solution.

\noindent \textbf{Bookmark} enables users to mark and store a solution in the solution gallery by clicking the bookmark button when hovering over a solution in either the solution window (either viewport) or the solution history.

\noindent \textbf{Rename} allows users to change the name of a solution.
The solution name can become a useful annotation allowing for organising and finding candidate solutions.

\section{User Study}
Our user study of the interactive optimisation tool was designed to evaluate: (1) Usability of the tool; (2) The \replaced{tool's}{tools} ability to support unforeseen scenarios (flexibility); (3) The \replaced{tool's}{tools} support for the \replaced{recommendations}{guidelines} and usefulness of these \replaced{recommendations}{guidelines} as evidenced by its use.

\subsection{Study Design}
\label{sec:study-design}
\noindent \textbf{Participants:} We recruited 10 participants\added{: 7 females and 3 males}.
They were PhD students from our university and employees from other organisations \added{with a bachelor's degree or higher}.
As end-users of interactive optimisation tool are more likely to be non-experts rather than optimisation experts, we only selected participants who did \emph{not} have computer science or optimisation expertise.
All participants had a normal or corrected-to-normal vision and were without any colour vision impairment.

\noindent \textbf{Procedure:} 
Participants used our interactive optimisation tool throughout the study.
The application domain presented to participants was that of a company that sends electricians (in trucks) to customers.
Each customer requires a certain fixed amount of time for the service, and this has a time window.
All trucks start from a central depot and return to the depot after servicing their customers. 

The study took roughly one hour and ten minutes on average. The study was run on a MacBook Pro notebook with a 2.6 GHz Intel i5 processor and a 13-inch screen.
It had three parts:

\vspace*{1mm}
\noindent
\emph{(a) Training}: This provided an introduction to the application domain, the solution and constraint representation, interface components (solution window, solution history and solution gallery) as well as user interactions
(\emph{Modify} a solution by re-assigning customers;
Add \& remove a customer \emph{lock} constraint;
Add \& remove a customer \emph{order} constraint;
\emph{Re-optimise} a solution when constraints have been modified;
\emph{Load} a solution to the solution window;
\emph{Bookmark} a solution to the solution gallery;
\emph{Rename} a solution in the solution gallery.)

    
\vspace*{1mm}
\noindent\emph{(b) Problem-solving}: Participants were then asked to solve an optimisation problem in four different scenarios.
The first scenario was the tool's intended use-case while the other three were designed to evaluate its flexibility in situations it wasn't explicitly designed to handle. 
At the end of each scenario, participants were asked to explain their strategies for solving the problem.
    
In \emph{Scenario 1} participants were asked to find a solution minimising the total travelling distance with no constraints apart from satisfying customer time-windows.
This was the tool's intended use-case.
In Part 1 of this scenario\added{, participants} could manually improve the solutions but re-optimisation was not allowed, then in Part 2 they were allowed to use re-optimisation.
This was to ensure that they experimented with both manual manipulation and re-optimisation.
Participants were given a maximum of 5 minutes for each part of the scenario.

\begin{figure}[tb]
 \centering
 \includegraphics[width=\linewidth, trim=0cm 6.6cm 21cm 0mm,clip=true]{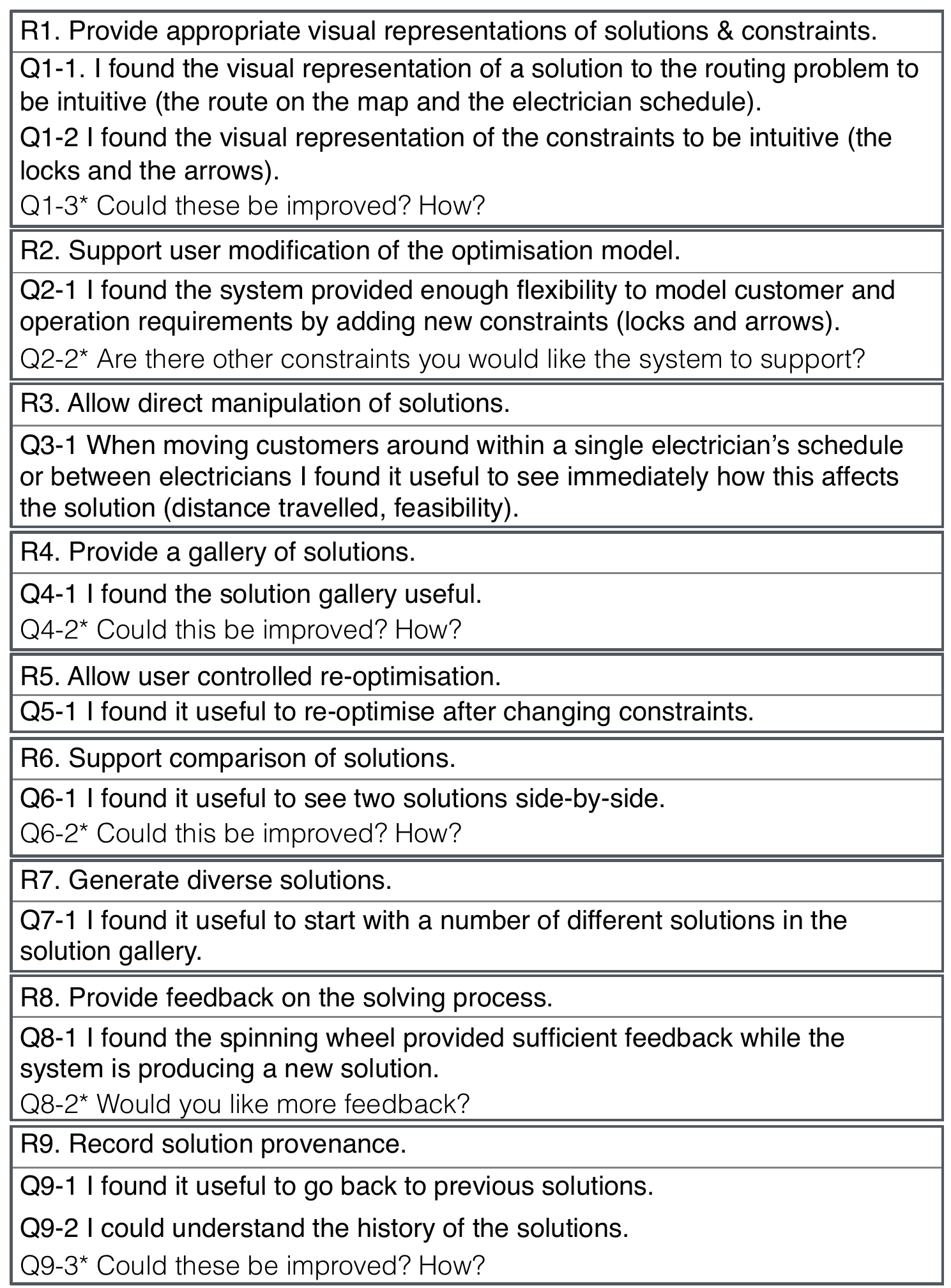}
 \vspace{-15pt}
 \caption{(\textbf{Q})uestions about interface features and their usefulness based on each \replaced{(\textbf{R})ecommendation}{(\textbf{G})uideline}. Questions marked with * are open-ended questions.}
 \vspace{-10pt}
 \label{fig:guideline-transform}
\end{figure}

\emph{Scenario 2} asked participants to solve a question of similar difficulty to Scenario 1.
However, we introduced another objective, asking participant's to minimise travel distance but also balance the workload of each truck so that they service close to the same number of customers.
There was a 5-minute time limit for this Scenario.
    
\emph{Scenario 3} asked participants to find a solution satisfying a list of customer requests.
The two objectives remained the same as in Scenario 2.
Most of the requests could be satisfied by adding either a customer lock constraint or a customer order constraint.
For instance, one request was ``Customer X requires the service from Electrician 2''.
However, there was a more difficult request specifying that one customer should not be serviced by a particular truck.
This was difficult because none of the two customer-related constraints can directly encode it.
Because of the difficulty, participants had 10 minutes for this scenario.
    
In \emph{Scenario 4}, participants were asked to revise the schedule as the result of an emergency in which one of the trucks has broken down.
The two objectives remained the same as in Scenario 2.
This was challenging because not only did participants need to re-assign the remaining customers of the broken truck but also participants had to ensure that all services delivered before the truck break-down time were unchanged, as these events were now in the past.
There were no constraints that could directly achieve this, which made this problem very difficult. For this reason, we gave participants 15 minutes.
    
\vspace*{1mm}
\noindent\emph{(c) System evaluation:} After completing the scenarios participants were asked to evaluate the system \emph{interface} and \emph{usability}. 
Participants were first asked to rank different aspects of the interface using a Likert scale from 1 to 5.
The questions were designed to evaluate the usefulness of these features and hence the usefulness of the underlying \replaced{recommendations}{guidelines}. 
We decided not to ask participants to evaluate the \replaced{recommendations}{guidelines} directly because the \replaced{recommendations}{guidelines} were rather abstract and difficult for non-experts to understand. 
The questions can be found in Fig\added{.}~\ref{fig:guideline-transform}.
We used the standard System Usability Scale (SUS)~\cite{brooke1996sus}, a 10-item questionnaire to measure the usability of our interface.
Again participants were asked to rank each item using the same 5-point Likert scale. 
We did not modify the questions in the SUS questionnaire.
However, we did provide extra context for the first and the last questions in order to clarify what was being asked.
For example, the first question was `\emph{I think that I would like to use this system frequently.}'
We provided the extra context \replaced{`}{'}\emph{Assume that you need to solve similar vehicle routing problems as a part of your job}.'

We prepared four problem instances of the VRPTW with similar complexity for the four scenarios (3 trucks, 12 customers).
For the first three scenarios, we pre-computed 3 diverse solutions that were stored in both the solution history and the solution gallery.
The last scenario, started with a single solution.
In terms of the solution quality, the first two problem instances contained \emph{poor} solutions which were 30\% worse than the best solution.
The last two problems contained \emph{best} or close-to-best (with 5\% solution quality margin) solutions.
We did this because we thought this would encourage participants to explore different solutions in the first two scenarios as finding an improved solution was relatively easy.

\subsection{Results \& Discussion}
Recall we wished to evaluate: (1) Usability of the tool; (2) Its flexibility to support unforeseen scenarios; (3) Support for the \replaced{recommendations}{guidelines} and their usefulness.

\begin{figure*}[t]
 \centering
 \includegraphics[width=\linewidth, trim=0cm 19.2cm 0cm 0mm, clip=true]{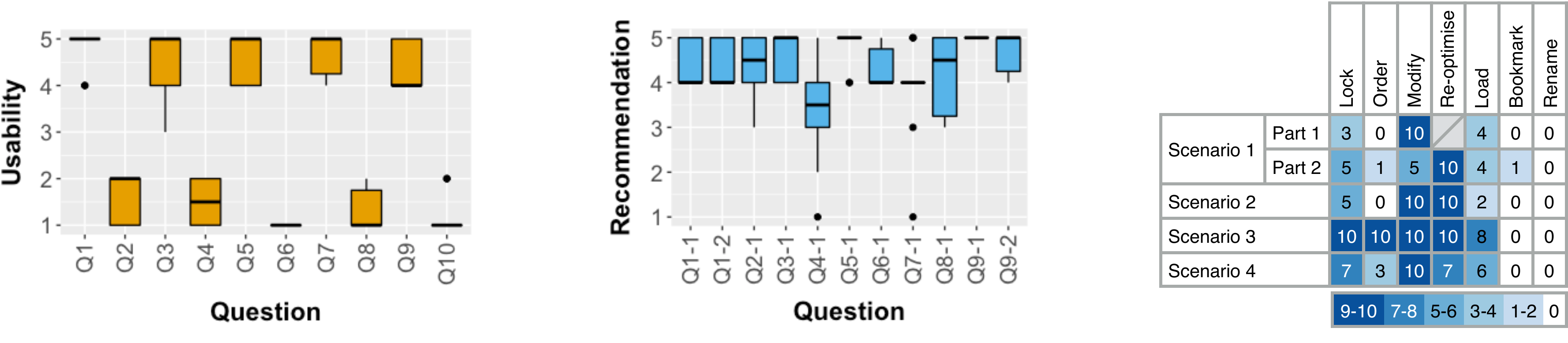}
 \vspace{-20pt}
 \caption{Participants' ratings of SUS questionnaire (left). Participants' ratings of interface questions (middle). Participants use of different kinds of interaction in each scenario (right).}
 \vspace{-10pt}
 \label{fig:user-study-figures}
\end{figure*}

\noindent \textbf{Usability:} 
At the end of the study participants were asked to evaluate the usability of the tool using the SUS (see Fig\added{.}~\ref{fig:user-study-figures} (left)).
We used the standard approach to calculate the overall system usability score:
convert the ratings to values from 0 to 4 (4 being the most positive response),
subtract the converted rating of each even-numbered questions from 4 because these questions were in a negative tone,
sum these converted ratings of all odd- and even-numbered questions,
then multiply the summed rating by 2.5 to give a score out of 100.

The highest score was 97.5, and the lowest score was 77.5.
Except for the lowest score, all other scores were equal to or greater to 87.5.
On average, the SUS score of our tool was 91, which is far better than the acceptance threshold value of 70~\cite{bangor2008empirical}.
Generally, a score above 80 means that the tool is of excellent usability, and is acceptable to end-users. Thus we can be reasonably confident that usability issues did not impact the use of different features.


\noindent \textbf{Flexibility:}
The different scenarios were designed to explore the tool's flexibility.
All participants finished both parts of Scenario 1.
8 out of 10 participants developed a better solution in Part 1.
However, none of them found an optimal solution, i.e. with the minimum total travel distance.
In general, participants tried to group close-by customers and service them with the same truck to reduce the total travelling distance.
In Part 2, an optimal solution was discovered by all participants using re-optimisation.

Most participants found very good solutions to Scenario 2.
These solutions were better than the starting solutions in two aspects: not only was the total distances shorter than any of the pre-computed diverse solutions, but also participants' developed solutions that were perfectly balanced with 4 customers per truck.
Having the second objective to also balance trucks' workloads affected participants' problem-solving strategies.

Specifically, 7 participants explained that they balanced the workload first by re-assigning customers, and then minimised the total travelling distance using re-optimisation.
In contrast, the other 3 participants tried to minimise the distance first using re-optimisation, and then to manually re-assign customers for workload balancing.
One participant had an interesting strategy.
(S)he primarily focused on re-assigning customers with long time windows because these customers were easier to fit in the schedule compared to short time-window customers.
Thus, such customer re-assignments were more tolerant of time window violations and it was easier to fix the solution when violations occurred.

Scenario 3 introduced customer requests.
All participants managed to find high quality solutions satisfying the requests and balancing the two objectives.
All participants added constraints to enforce the majority of the customer requests and then tried to balance the two objectives using the same approaches as in Scenario 2.
Most used lock to ensure that the customer who did not like a particular electrician remained assigned to one of the other electricians.


Scenario 4 asked participants modify the schedule after a truck breakdown.
9 out of 10 participants developed a valid solution.  
All participants agreed that the emergency should be handled first, i.e. first find a solution that no longer used the broken truck but that preserved the schedule for the time before the breakdown.
To do so they re-assigned customers from the broken truck to the other two trucks.
When doing so they also attempted to balance the workloads.
They then attempted to use re-optimisation to reduce the overall travel distance.
However, this typically produced an invalid solution as the optimisation solver could assign a customer back to the broken truck if the customer was unlocked.

After realising this, some participants decided to lock all 12 customers to prevent unwanted customer re-assignments before re-optimising.
This only worked for one fortunate participant.
The other participants found that the optimisation solver could not re-optimise the current solution due to conflicting constraints.

As a result, most participants developed their valid solutions through an iterative process, in which they re-assigned customers first before utilising re-optimisation to improve the travelling distance.
Once re-optimisation was done, they re-assigning customers before the next round of re-optimisation.
This process terminated when a valid solution was found.
However, three participants solved the problem using a purely manual approach to re-assign customers without relying on re-optimisation to find a solution.


Thus we see that, as hoped, the tool allowed the participants to flexibly handle a variety of scenarios which the tool was not explicitly designed to support.
This relied upon the ability to add/modify/remove customer constraints, re-optimise taking into account modified constraints, and to manually modify a solution.

\noindent \textbf{Support for the \replaced{recommendations}{guidelines} and their usefulness:} 
Fig.~\ref{fig:user-study-figures} (right) shows the number of participants who used various features of the interface in each scenario and Fig.~\ref{fig:user-study-figures} (middle) summarises participants response to the questions about the system interface.
These, together with the participant strategies detailed above, allow us to gauge how well the tool supported each \replaced{recommendation}{guideline} and at least to some extent the usefulness/importance of the underlying \replaced{recommendations}{guidelines}. \added{We first detail the recommendations with strong support.}


\emph{\replaced{Recommendation}{Guideline} 1}: This was well supported by the tool.
Most participants gave 4 out of 5 for the visual representations of the solution (Q1-1) and constraints (Q1-2) (see Fig.~\ref{fig:guideline-transform}).
Participants agreed that both the schedule and street map were clean and well designed.
One participant said ``\emph{The schedule looks very straightforward, and from the map, I can clearly see customers as well as their [assigned trucks].}''
Another commented ``\emph{It is very easy to understand what is happening.}'' implicitly suggesting that the visualisation is useful.
Participants did suggest two improvements (Q1-3): to add route arrows on the street map to indicate travel direction, and to place the customer lock constraint label in front of the customer letter.
While participant comments did not directly address the importance of this \replaced{recommendation}{guideline}, understandable solution and constraint representation underlie \replaced{Recommendations}{Guidelines} 1-4 and 6, most of which we found to be important.



\emph{\replaced{Recommendation}{Guideline} 2}: This was also well supported by the tool.
On average, participants ranked 4.5 for flexibility (Q2-1). Our observations of participants suggested they understand both customer lock and order constraints, and they could effectively use these to model more complex requirements.
From Fig.~\ref{fig:user-study-figures} (right) we see that they were used by at least half of the participants in each scenario except for Scenario 1 Part 1, and by all in Scenario 3.
This argues strongly for the importance of this \replaced{recommendation}{guideline} to provide flexibility in unforeseen situations.



When participants were asked if there were other constraints they would like the system to support (Q2-2), some suggested that it would be nice to have a truck lock constraint which locks all customers serviced by the same truck and also prevents other customers being assigned to this particular truck during re-optimisation.
Some also suggested that apart from the existing ``relative'' customer lock which only guarantees a customer to be serviced by a specific truck, it would be nice to have an ``absolute'' customer lock so that once a customer is locked, the service order of this particular customer is also locked.

\emph{\replaced{Recommendation}{Guideline} 3:} This was very well supported.
The majority of the participants ranked 5 for solution manipulations and immediate feedback after manipulations (Q3-1).
All participants agreed that it was necessary to provide such immediate feedback when changes are made.
One participant commented ``\emph{I think it (manipulating a solution and providing feedback) is very useful. Because not only I can see how distance (one of the objectives) has changed, but also I can see whether the solution has any violations highlighted in red.}''
Manual adjustment of solutions was performed by all participants in all scenarios, strongly suggesting the importance of this \replaced{recommendation}{guideline}.

\deleted{Guideline 4: We were surprised to find that participants did not find the solution gallery to be useful with an average ranking of 3 (Q4-1).
Only one participant used the \emph{bookmark} interaction, and none used the \emph{rename} interaction though
five participants used the \emph{load} interaction.}

\deleted{
However, it seems that one reason participants did not use the solution gallery was the enforced time limit in each scenario. We observed that some participants utilised the maximum amount of time in several scenarios and one participant stated that
``If you do not put the time limit [for each scenario], it (the solution gallery) would be useful. With the time limit, I have to hurry up and focus on only one solution.''
When asked `What if there was no time limit, then how would you use it?' they responded ``I would like to store all good solutions in it (the solution gallery) and do comparisons afterwards.''}

\deleted{
An associated reason for not using the solution gallery was that the number of solutions is relatively small.
One participant said ``Actually I only deal with one solution at a time. But if there are a lot of solutions, then maybe that (the solution gallery) helps.''
Another one responded ``It might be more useful when the problem becomes more complex with more solutions provided.''
Another reason may be related to lack of experience with the tool.
One participant commented ``Maybe if I use this software more, I would have more experience [using the solution gallery].''
}
    
\deleted{
When participants were asked how to improve the solution gallery (Q4-2), one suggested ``If there is a way to [automatically] re-order solutions based on quality, it would be great, for instance, to place the best solution at the very top and the worst at the very bottom.''
}

\deleted{
We also conjecture that a solution gallery would be more useful in more complex multi-criteria optimisation and that it should support ranking based on multiple criteria.
This was supported by the suggestions of two participants who mentioned customer distributions within trucks.
Specifically, one said ``It would be better to include the workload balance (customer distribution) information of each electrician [for each solution in the solution gallery].''
}

\emph{\replaced{Recommendation}{Guideline} 5:} 
This was very well supported with all but one participant finding re-optimisation very useful (Q5-1) and the other finding it useful.
One participant commented ``\emph{This is extremely useful because I feel like I do not need to modify the solution again after re-optimisation.}''
Re-optimisation was used by all participants in three scenarios and by most in Scenario 4.

\added{Some participants suggested that it would be nice to clearly see the changes between solutions.
Two participants advised demonstrating the solution changes before and after re-optimisation with one saying
``\emph{After re-optimisation, the orders of customers may have changed drastically. Therefore it would be great to show what and where the changes are.}''  One participant commented ``\emph{If it can show animations to indicate how does one solution transform into another, that would be useful because it could reduce my memory load and remind me what has changed.}'' This should be explored in future work.}

\deleted{\emph{Guideline 6:} This was supported with all participants agreeing that seeing two solutions side-by-side was useful (Q6-1).
However, in practice we did not observe participants spending a lot of time comparing solutions, instead they appeared to focus on one solution and try and improve it.
As for \emph{Guideline 4} it is likely that this was because of the time limit.
In particular, one participant said ``Without time limitation, I would like to develop two solutions with one solution per viewport and choose the better one in the end.''
Another participant made a similar comment.}
    
\deleted{
Considering participants' suggestions on the side-by-side layout (Q6-2), more than half of the participants thought having two solutions presented simultaneously would be sufficient.
For instance, one participant commented ``I think two is enough. With more than two solutions, I won't be able to simultaneously look at them all.''
However, one participant said ``If you can provide options [of how many solutions to be presented], depending on the size of monitors, they can make it as many [solutions] as they want to. But I think two is enough.''}

\deleted{
\emph{Guideline 7:} This was supported with most participants agreeing that it is beneficial to have several different solutions in the solution gallery, to choose from during the problem-solving (Q7-1).
In Scenario 3, eight participants conducted a quick examination of all solutions in the gallery to try to find the solution satisfying the greatest number of customer requests.
They then used this as their starting point.
Thus, there is reasonable support for the usefulness of providing a gallery of diverse solutions.}

\deleted{
However, our observations reveal that after an initial examination, most participants focused on one solution and did not switch to another solution.
Again this is probably because of the time limitation, as for \emph{Guideline 4} and \emph{6}.
However, in Scenarios 4, we did see three participants switch to another solution when they became stuck.
The situation improved after switching, resulting in better solution performance.
}


    
\emph{\replaced{Recommendation}{Guideline} 8:} There was \replaced{qualified}{some} support for this.
Most participants believed having the spinning wheel during re-optimisation provided sufficient feedback about the ongoing solving process (Q8-1).
In particular, one participant commented ``\emph{It is a must-have. It is rather a psychological effect telling me that the re-optimisation is still running.}''
Another three participants made similar comments.
However, three participants thought the spinning wheel was not useful.
This is because they were too used to it.
One participant said ``\emph{It is so common that every system has this (a spinning wheel or something similar to indicate task progress).}''

When participants were asked whether more feedback is needed (Q8-2), several suggestions were provided.
Specifically, one participant mentioned a progress bar and commented ``\emph{I think it (the spinning wheel) is enough since it (the re-optimisation) is fast. If it (the re-optimisation) is slow, you might want to consider something like a progress bar.}''
Another participant suggested having a pop-up message to briefly summarise all objective values.

\added{Overall, this suggests that while providing feedback on the solver's progress is generally appreciated, more research is needed into how to do so. One must be careful not to provide feedback that leads to over trust~\cite{liusubmitted}.}

\emph{\replaced{Recommendation}{Guideline} 9} was very well supported.
Every participant believed that it was essential to be able to go back to previous solutions (Q9-1).
For instance, one participant said ``\emph{It is really useful because you may make a mistake, and you may also not remember the previous solution.}''
All participants also agreed that the history of the solutions is easy to understand (Q9-2).
Participant's behaviour also showed that it was relatively common to load a solution from the history in many scenarios (see Fig.~\ref{fig:user-study-figures} (right)).

\deleted{Two possible improvements were suggested by participants (Q9-3).
One participant suggested to include a workload chart in the history.
(S)he said ``I think the chart (solution histogram) is really useful because I can see all my solutions.
However, it only shows [the overall travelling] distance. If I would like to see the workload balance between solutions, it would be nice to have a similar chart for the workload so that I could have a complete history [of all solutions].''
Second, similar to the above analysis of {Guideline} 8 (Q8-2), participants suggested that it would be nice to clearly see the changes between solutions.
In particular, one participant commented ``If it can show animations to indicate how does one solution transform into another, that would be useful because it could reduce my memory load and remind me what has changed.''}

\added{\noindent\textbf{Recommendations involving multiple solutions received less support}. Recommendations 4, 6 and 7 were generally not as well supported by participant feedback, as follows.

\emph{Recommendation 4:} In particular, we were surprised to find that participants did not find the solution gallery to be useful with an average ranking of 3 (Q4-1).
Only one participant used the \emph{bookmark} interaction, and none used the \emph{rename} interaction though
five participants used the \emph{load} interaction. When participants were asked how to improve the solution gallery (Q4-2), one suggested ``\emph{If there is a way to [automatically] re-order solutions based on quality, it would be great, for instance, to place the best solution at the very top and the worst at the very bottom.}''

\emph{Recommendation 6:} While this was supported with all participants agreeing that seeing two solutions side-by-side was useful (Q6-1)\replaced{.}{,} \replaced{In}{in} practice we did not observe participants spending a lot of time comparing solutions\replaced{.}{,} \replaced{Instead}{instead} they appeared to focus on one solution and try and improve it. Considering participants' suggestions on the side-by-side layout (Q6-2), more than half of the participants thought having two solutions presented simultaneously would be sufficient.
For instance, one participant commented ``\emph{I think two is enough. With more than two solutions, I won't be able to simultaneously look at them all.}''
However, one participant said ``\emph{If you can provide options [of how many solutions to be presented], depending on the size of monitors, they can make it as many [solutions] as they want to. But I think two is enough.}''

\emph{Recommendation 7:} This had more support with most participants agreeing that it is beneficial to have several different solutions in the solution gallery to choose from (Q7-1).
In Scenario 3, eight participants conducted a quick examination of all solutions in the gallery to try to find the solution satisfying the greatest number of customer requests.
They then used this as their starting point.
However, our observations revealed that after an initial examination, most participants focused on one solution and did not switch to another solution. Nonetheless, in Scenarios 4, we did see three participants switch to another solution when they became stuck.
The situation improved after switching, resulting in better solution performance.

We suspect that one reason that many participants did not use the gallery or explore multiple solutions was the enforced time limit in each scenario. We observed that some participants utilised the maximum amount of time in several scenarios and one participant stated that
``\emph{If you do not put the time limit [for each scenario], it (the solution gallery) would be useful. With the time limit, I have to hurry up and focus on only one solution.}''
When asked `What if there was no time limit, then how would you use it?' they responded ``\emph{I would like to store all good solutions in it (the solution gallery) and do comparisons afterwards.}''
Another participant said ``\emph{Without time limitation, I would like to develop two solutions with one solution per viewport and choose the better one in the end.}''
A further participant made a similar comment.

Lack of experience with the tool may also have contributed.
One participant commenting ``\emph{Maybe if I use this software more, I would have more experience [using the solution gallery].}''
    
Another part of the reason is that participants felt the problems were simple enough that they did not feel the need to explore multiple solutions.
One  said ``\emph{Actually I only deal with one solution at a time. But if there are a lot of solutions, then maybe that (the solution gallery) helps.}''
Another one responded ``\emph{It might be more useful when the problem becomes more complex with more solutions provided.}''

We conjecture that a solution gallery seeded with diverse solutions and the ability to compare solutions would be more useful in more open-ended tasks such as design involving complex multi-criteria optimisation. This would accord with Dayama et al.~\cite{dayama2020grids} who found designers liked to have a gallery of automatically suggested solutions.  Such a gallery  should support ranking based on multiple criteria.
This was supported by the suggestions of two participants who mentioned customer distributions within trucks.
Specifically, one said ``\emph{It would be better to include the workload balance (customer distribution) information of each electrician [for each solution in the solution gallery].}''
}

\section{Conclusion}
We have presented a comprehensive set of design \replaced{recommendations}{guidelines} (\replaced{R}{G}1-9) for interactive optimisation systems.
To the best of our knowledge these are the first such \replaced{recommendations}{guidelines}.
An examination of 15 representative systems from the literature revealed that none supported all of the \replaced{recommendations}{guidelines}.
Some \replaced{recommendations}{guidelines}--appropriate visual representation and modifiable constraints--were well supported, some were supported by about half of the systems--manual modification of a solution, user controlled re-optimisation and generation of diverse solutions--while the remaining \replaced{recommendations}{guidelines}--solution gallery, side-by-side solution comparison, solver feedback and record of solution provenance--were rarely supported.

We then evaluated the \replaced{recommendations}{guidelines} using the vehicle routing problem with time windows (VRPTW) as an exemplar application.
We built an interactive tool for solving these problems that was informed by the \replaced{recommendations}{guidelines}.
Ten participants then used this system to solve a variety of routing problems.
These showed that the tool was very flexible and could be used to solve variants of VRPTW it was not originally designed for.
\replaced{Participants found the tool highly usable and, based on user feedback and behaviour, we found strong support for most of the underpinning recommendations.}{Participants found the tool highly usable and generally supported the design of all its features.
Based on their feedback and observed behaviour we found strong support for most of the guidelines.} 

\replaced{The recommendations supporting exploration of  multiple solutions (R4 and R6) received less support in our observations. We conjecture that  this is likely to reflect the limited time allowed in the study; the simplicity of the objective function; and the nature of the tasks they were required to perform.  Further studies are required to investigate this, ideally with longitudinal deployment of interactive optimisation into a problem-solving workflow.}{While there was less support for the solution gallery and side-by-side comparison, this is likely to reflect the limited time allowed in the study.}  

Of course\added{, in general,} more case studies are required in order to confirm the general applicability of the \replaced{recommendations}{guidelines}.
Nonetheless \replaced{we believe that our}{our initial review of the literature and study suggests that the} \replaced{recommendations}{guidelines} are a valuable addition to our understanding of how to design interactive optimisation systems as the features they suggest are useful but many are not commonly found in existing systems.

\acknowledgments{
We thank Agilent Technologies Inc. for their support through a Thought Leader Award and the support of Data 61, CSIRO which is funded by the Australian Government through the Department of Communications and the Australian Research Council through the ICT Centre for Excellence Program. \added{We also thank the reviewers for their insightful comments and suggestions.}
}

\bibliographystyle{abbrv-doi}

\bibliography{references}
\end{document}